\documentclass{emulateapj}

\newcommand{\etal}{{\it et al.}}

\begin{document}

\title{What Lies Beneath: Using p(z) to Reduce Systematic Photometric Redshift Errors}

\shorttitle{}

\author{D. Wittman \\
Physics Department, University of California, Davis,
  CA 95616; dwittman@physics.ucdavis.edu}


\begin{abstract} 
We use simulations to demonstrate that photometric redshift ``errors''
can be greatly reduced by using the photometric redshift probability
distribution $p(z)$ rather than a one-point estimate such as the most
likely redshift.  In principle this involves tracking a large array of
numbers rather than a single number for each galaxy.  We introduce a
very simple estimator that requires tracking only a single number for
each galaxy, while retaining the systematic-error-reducing properties
of using the full $p(z)$ and requiring only very minor modifications
to existing photometric redshift codes.  We find that using this
redshift estimator (or using the full $p(z)$) can substantially reduce
systematics in dark energy parameter estimation from weak lensing, at
no cost to the survey.
\end{abstract}

\section{Introduction}

Photometric redshifts (Connolly \etal\ 1995, Hogg \etal\ 1998, Benitez
2000) are a key component of galaxy surveys.  As surveys get larger,
reducing statistical uncertainties, systematic errors become more
important.  Systematic errors in photometric redshifts are therefore a
top concern for future large galaxy surveys, for example as
highlighted by the Dark Energy Task Force (Albrecht \etal\ 2006).

Much of the concern has centered on ``catastrophic outliers'' which
are galaxies for which the photometric redshift is very wrong, for
example when mistaking the Lyman break at $z\sim 3$ for the
4000\AA\ break at very low redshift.  Even a small fraction of
outliers can significantly impact the downstream science, and modeling
this impact requires going beyond simple Gaussian models of
photometric redshift errors.

In many cases, however, outliers are ``catastrophic'' only because
they have a multimodal redshift probability $p(z)$, which cannot be
accurately represented by a single number such as the most probable
redshift.  Fernandez-Soto \etal\ (2002) showed that after defining
confidence intervals around the $p(z)$ peaks, 95\% of galaxies in
their sample had spectroscopic redshifts within the 95\% confidence
interval and 99\% had spectroscopic redshifts within the 99\%
confidence interval.  Yet the same data appear to contain catastrophic
outliers on a plot where each galaxy is represented only by a point
with symmetric errorbars.

There is a second motivation for using the full $p(z)$.  The redshift
ambiguities described above are due to color-space degeneracies.  But
even without these degeneracies, photometric redshift errors should be
asymmetric about the most probable redshift due to the nonlinear mapping
of redshift into color space.  Avoiding biases from this effect also
requires reference to $p(z)$.  Indeed, Mandelbaum \etal\ (2008) showed
that using $p(z)$ in Sloan Digital Sky Survey (SDSS) data (which are
not deep enough to suffer serious degeneracies) substantially reduced
systematic calibration errors for galaxy-galaxy weak lensing.

In this paper we demonstrate, using simple simulations, the reduction
in systematic error that can result from using the full $p(z)$ in a
deep survey with significant degeneracies.  We also introduce a simple
way to reduce the computational cost of doing so.

\section{Simulations}

We conducted simulations similar to those in Margoniner \& Wittman
(2008) and Wittman \etal\ (2007), which used the Bayesian Photometric
Redshift (BPZ, Benitez 2000) code, including its set of six template
galaxy spectral energy distributions (SEDs) and its set of priors on
the joint magnitude-SED type-redshift distribution.  We started with
an actual R band catalog from the Deep Lens Survey (DLS, Wittman
\etal\ 2002).  For each galaxy, we used the $R$ magnitude to generate
a mock type and redshift according to the priors, and then generated
synthetic colors in the BVRz$^\prime$ filter set used by DLS.  (The
filter set is not central to the argument here, but one must be used
for concreteness.)  We then added photometry noise and zero-point
errors representative of the DLS.  The color distributions in the
resulting mock catalog were similar to those of the actual catalog,
indicating that the mock catalog is consistent with a real galaxy
survey.

We then ran the mock catalog of 83,000 galaxies through BPZ, saving
the full $p(z)$.  In a post-processing stage, we can extract from the
full $p(z)$ not only the most probable redshift (which had already
been determined by BPZ and labeled $z_B$), but other candidate
one-point estimates such as the mean and median of $p(z)$, as well as
the summed $p(z)$ for any desired set of galaxies.

\section{Results}

\begin{figure}
\centerline{
\resizebox{3.5in}{!}{\includegraphics{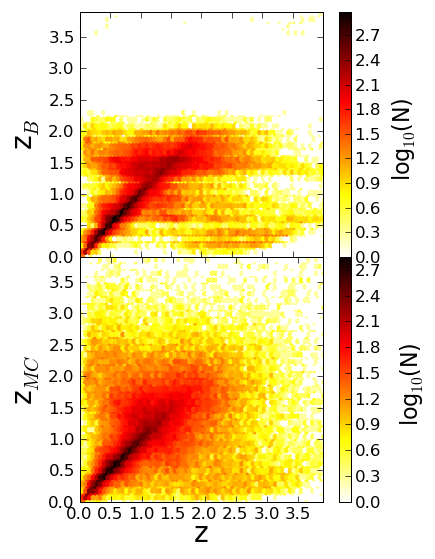}}
}
\caption{Top: each galaxy's most probable photometric redshift ($z_B$)
  vs. spectroscopic redshift.  Bottom: Monte Carlo sampling from each
  galaxy's redshift probability distribution ($z_{MC}$)
  vs. spectroscopic redshift.  Sampling the distribution cleans up
  many artifacts introduced by considering only the most probable
  photometric redshift.
\label{fig-zmc}}
\end{figure}

We first show one of the traditional one-point estimates to more
clearly illustrate the problem.  The left panel of
Figure~\ref{fig-zmc} shows the most probable redshift $z_B$ vs. true
redshift $z$.  To accurately render both the high- and low-density
parts of this plot, we show it as a colormap rather than a
scatterplot.  The core is rather tight, requiring a logarithmic
mapping between color and density to bring out the more subtle
features in the wings.  With this mapping, the systematics are clear:
a tendency to put galaxies truly at $z\sim 2-3$ at very low redshift;
a tendency to put galaxies truly at low redshift at $z_B\sim 1.4-2$;
and asymmetric horizontal smearing in several different $z_B$
intervals, e.g. at $z_B\sim 1.5$.  The specifics of the features
depend on the filter set, but their general appearance is typical (but
note that they will be difficult to see in plots based on
spectroscopic followup of deep imaging surveys, as the brighter,
spectroscopically accessible, galaxies form a much tighter relation).
Corresponding plots are not shown for other one-point estimates such
as the mean and median of $p(z)$, but they have similar to worse
systematic deviations.

Which systematics are most important depends on the application. In
this paper we consider two-point correlations of weak gravitational
lensing (cosmic shear), which require that the photometric redshift
{\it distribution} of a sample of galaxies be as close as possible to
the true redshift distribution; errors on specific galaxies are not
important.  Also, accurate knowledge of the scatter is much more
important than minimizing the amount of scatter. Thus,
Figure~\ref{fig-zmc} would ideally be symmetric about a line of unity
slope, reflecting the fact that the photometric and true redshift
distributions are identical.\footnote{In practice, the two
  distributions would be the same for any subsample, not only for the
  entire sample.  We address this later, in Figure~\ref{fig-sumpz}.}
This is clearly not the case for the top panel of
Figure~\ref{fig-zmc}.

\begin{figure}
\centerline{
\resizebox{3.5in}{!}{\includegraphics{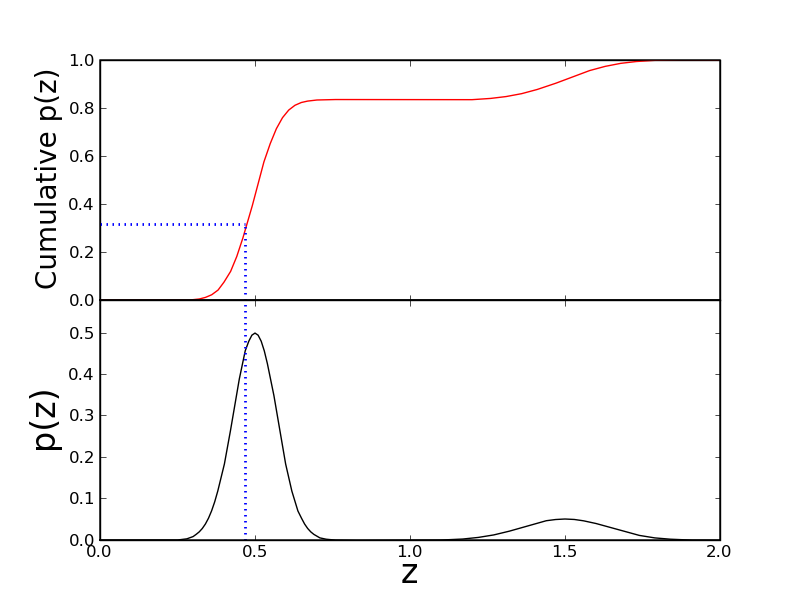}}
}
\caption{Construction of the $z_{MC}$ estimator for an example
  galaxy. The bottom panel shows $p(z)$ as usually plotted.  The top
  panel shows the cumulative $p(z)$, that is, the probability that the
  galaxy lies at redshift less than the value on the abscissa.  A
  random number in the range 0-1 is drawn, in this case 0.32, and the
  redshift at which the cumulative $p(z)$ has a value of 0.32 is
  recorded as the Monte Carlo estimate $z_{MC}$.
\label{fig-mcprocess}}
\end{figure}

\begin{figure}
\centerline{\resizebox{3.75in}{!}{\includegraphics{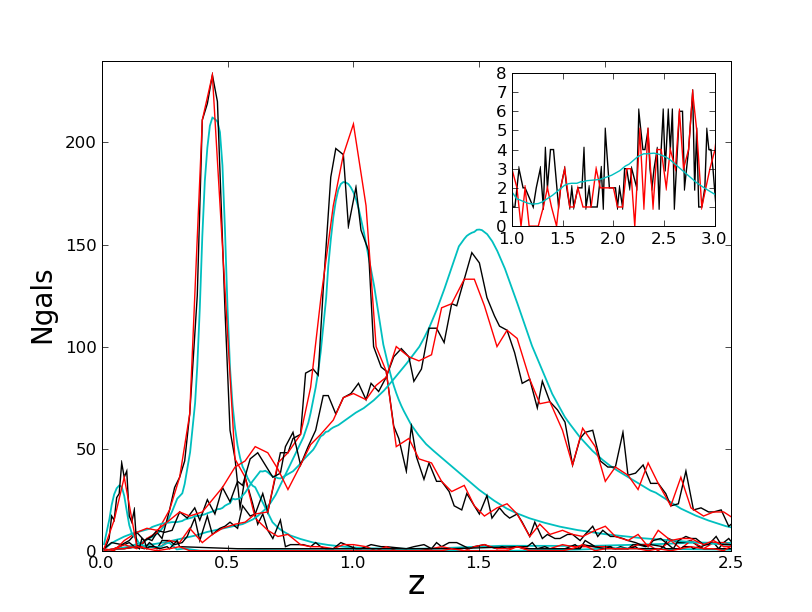}}}
\caption{Black: true redshift distributions of photometric redshift
  cuts $z_B<0.1$, 0.4-0.5, 0.9-1.1, and 1.4-1.6 in the simulation.
  Red: distributions of the $z_{MC}$ estimator for the same sets of
  galaxies. Blue: summed $p(z)$ for the same sets of galaxies.  The
  simple estimator $z_{MC}$ does as well as tracking the full $p(z)$.
  Using each galaxy's most probable redshift $z_B$ to infer the redshift
  distributions would result in four vertical-sided bins.  Inset: a
  high-redshift bump in the true redshift distribution of galaxies
  with $0.4<z_B<0.5$ is captured by the summed $p(z)$ or by $z_{MC}$,
  but would be considered ``catastrophic outliers'' if $z_B$ were
  used.
\label{fig-sumpz}}
\end{figure}

Ideally $p(z)$ contains the required information lacking in the single
number $z_B$, but it is also more difficult to work with, requiring
the storage and manipulation of an array of numbers for each galaxy.
We simplify the computational bookkeeping by defining a single number
which is by construction representative of the full $p(z)$.  This
estimate is simply a random number distributed according to the
probability distribution $p(z)$ and is denoted by $z_{MC}$ because it
is a Monte Carlo sample of the full $p(z)$.  Specifically, for each
galaxy, a random number $x$ is drawn uniformly from the interval
$[0,1)$, and the Monte Carlo redshift $z_{MC}$ is defined such that
  $\int_0^{z_{MC}} p(z^\prime)dz^\prime = x$.
  Figure~\ref{fig-mcprocess} illustrates the process.  The bottom
  panel shows $p(z)$ as usually plotted, while the top panel shows the
  cumulative $p(z)$, that is, the probability that the galaxy lies at
  redshift less than the value on the abscissa.  A random number in
  the range 0-1 is drawn, in this case 0.32, and the redshift at which
  the cumulative $p(z)$ has a value of 0.32 (dotted line) is recorded
  as $z_{MC}$, in this case 0.47.  This results in a single number for
  each galaxy, which remains unbiased even if $p(z)$ is multimodal
  and/or asymmetric.  Of course, some precision is lost in this
  process; it should be avoided when studying a small number of
  galaxies in great detail, but for large samples of galaxies it must
  converge to the $p(z)$ of the sample.  Furthermore, it requires only
  a minor modification to most photometric redshift codes.

The bottom panel of Figure~\ref{fig-zmc} shows $z_{MC}$ vs. $z_B$.
Clearly, the systematics are vastly improved.  Even with the
logarithmic scaling, it is difficult to see departures from symmetry
about a line of unity slope. We therefore compare one-dimensional
histograms in what follows.  
A typical use of photometric redshifts in a galaxy survey will be to
bin the galaxies by redshift, for example to compute shear
correlations in redshift shells.  Because the true redshifts will not
be known, the galaxies must be binned by some photometric redshift
criterion.  For simplicity, we choose $z_B$.

Figure~\ref{fig-sumpz} (upper panel) shows true and inferred redshift
distributions of galaxies in four $z_B$ bins: 0-0.1, 0.4-0.5, 0.9-1.1,
and 1.4-1.6.  The true distributions is shown in black, the
distribution inferred from $z_{MC}$ is shown in red, and the
distribution inferred from summing the galaxies' $p(z)$ is shown in
blue.  The asymmetry and the wings of the true redshift distribution
are well captured by $z_{MC}$ or by summing $p(z)$.  By comparison,
using each galaxy's most probable redshift $z_B$ to infer the redshift
distributions would have resulted in four vertical-sided bins, which
would become roughly Gaussian after convolving with the typical
galaxy's $z_B$ uncertainty.  It is clear that this would not capture
the true redshift distribution nearly as well as the $p(z)$ method
does.  For example, the inset in Figure~\ref{fig-sumpz} shows a small
high-redshift ($z\sim 2.5$) bump in the 0.4-0.5 bin which is captured
by $z_{MC}$ or by summing $p(z)$.  Looking only at the most probable
redshift $z_B$ would result in these galaxies being considered
catastrophic outliers.

Another way to reduce ``catastrophic outliers'' and related
systematics might be to discard galaxies whose $p(z)$ is multimodal,
not sharply peaked, or otherwise fails some test.  This may be
effective, but it greatly reduces the number of galaxies available to
work with.  The $z_{MC}$ and full $p(z)$ methods accurately reflect
the true redshift distributions without requiring any reduction in
galaxy sample size.  This is important for applications such as
lensing, for which galaxy shot noise will always be an issue.

Having demonstrated that using $p(z)$ (whether by sampling or by using
the full distribution) greatly reduces photometric redshift
systematics, two questions naturally arise.  How much better is it in
terms of a science-based metric?  And what are the remaining errors or
limitations?  Because $z_{MC}$ and the full $p(z)$ give very similar
results, references to using $p(z)$ in the remainder of the paper
should be understood to include either implementation.

\section{Effect on Dark Energy Parameter Estimation}

\begin{figure}
\centerline{\resizebox{3.5in}{!}{\includegraphics{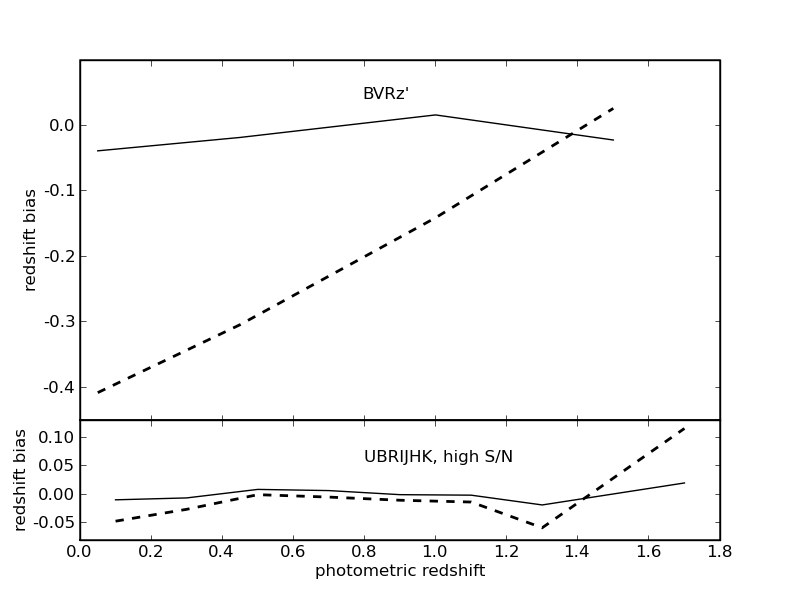}}}
\caption{Redshift bias (inferred minus true mean redshift) for the
  four bins shown in Fig.~\ref{fig-sumpz}, when using the full $p(z)$
  (solid) and when using the most probable redshift $z_B$ for each
  galaxy (dotted).  The results for $z_{MC}$ are nearly
  indistinguishable from those for the full $p(z)$ and are not shown.
  {\it Top panel:} BVRz$^\prime$ filter set and DLS noise model.  {\it
    Bottom panel:} UBRIJHK filter set and optimistic noise model with
  photometric signal-to-noise of 10 in {\it each} filter, regardless
  of redshift or magnitude.
\label{fig-zbias}}
\end{figure}

For each of the four $z_B$ bins shown in Fig.~\ref{fig-sumpz}, we
compute the redshift bias (inferred minus true mean redshift) for the
full $p(z)$ and $z_B$ approaches.  These are shown in
Fig.~\ref{fig-zbias} as the solid and dotted lines respectively (the
results for $z_{MC}$ are nearly indistinguishable from those for the
full $p(z)$ and are not shown).  For the DLS filter set and noise
model, the bias is within a few hundredths of a unit redshift when
using $p(z)$, but only within several tenths when using the most
probable redshift $z_B$.  For lensing applications, the $z_B$ bias at
low redshift is exaggerated somewhat, because it results from a small
outlying bump at high redshift, as shown in the inset of
Fig.~\ref{fig-sumpz}.  In a real survey, these high-redshift
interlopers would be smaller and fainter than the galaxies that really
belong in the low-redshift bin, and therefore have less precise shape
measurements and smaller weight on the shear statistics.  We account
for this effect by assigning a lensing weight to each mock galaxy
based on its magnitude and drawn from the actual distribution of
weights as a function of magnitude in the DLS.  This reduces the bias
of the lowest-redshift bin to 0.02 for $p(z)$ and 0.24 for $z_B$, and
has a progressively smaller effect on higher-redshift bins.  

Fig. 7 of Ma \etal\ (2006) shows the degradation in dark energy
parameters for a wide and very deep weak lensing survey, as a function
of the tightness of priors that can be put on the redshift bias and
scatter in photometric redshift bins.  It shows that loose priors of
order 0.2 result in 80-85\% degradation in $w_0$ estimates (with
respect to a survey with absolutely no redshift errors).  If, on the
other hand, one need only allow for a $\sim0.02$ bias as with the
$p(z)$ approach, the degradation decreases to 50-60\%.  For estimating
$w_a$, the degradation decreases from a factor of six to a factor of
about 2.5 by employing $p(z)$.

These are only rough estimates, for a number of reasons.  Future
surveys as deep as those contemplated by Ma \etal\ (2006) will use
more extensive filter sets, which will probably improve the $z_B$
performance somewhat with respect to the $p(z)$ performance.  And
nearer-term, shallower surveys have looser redshift requirements
because their shear measurements are not as precise.  But it is clear
that using $p(z)$ greatly improves the survey at essentially no cost.
We also conducted simulations using a more extensive filter set, to
check the generality

\section{Remaining Errors and Limits of this Work}

The simulations are simplistic in that the same six SED templates (and
priors) used to infer $p(z)$ are used to generate the mock catalogs.
In real photometric redshift catalogs, $p(z)$ will be less perfect
because galaxy SEDs are more varied and priors are imperfectly known.
Smaller effects of the same nature include uncertainties in real-life
filter and throughput curves, which are artificially reduced to zero
here.  However, these errors also affect the most probable redshift.
Therefore, although the simulated results presented here are
optimistic overall (given this filter set), the performance of $p(z)$
{\it relative} to using the most probable redshift may not be.  More
sophisticated simulations beyond the scope of this paper will be
required to determine the limits of $p(z)$ accuracy for any given
filter set and survey depth.

The remaining redshift bias is not trivial, 0.01-0.02.  This is an
order of magnitude larger than required to keep the $w_0$ degradation
within 10\% of an ideal survey (for the deepest surveys; requirements
are less stringent for shallower surveys).  However, a few factors are
actually pessimistic here compared to future large surveys: the
limited filter set, and relatively large zeropoint errors ($\sim 0.03$
mag here vs 0.01 mag for SDSS and future large surveys).  A more
extensive filter set will improve {\it both} the $p(z)$ and the $z_B$
etimates, but will probably improve the $z_B$ estimate more as it
eliminates some degeneracies.  Again, survey-specific simulations will
be required to make more specific conclusions.

Finally, there is a source of bias not simulated here: Eddington
(1913) bias.  The type and redshift priors are based on magnitude, but
at the faint end magnitudes are biased due to the asymmetry between
the large number of faint galaxies that noise can scatter to brighter
magnitudes, versus the smaller number of moderately bright galaxies
that noise can scatter to fainter magnitudes.  Surveys wishing to
derive photometric redshifts for galaxies detected at, say, 10$\sigma$
or fainter, must use the Hogg \& Turner (1998) prescription for
removing Eddington bias from each galaxy's flux measurements if they
are to avoid nontrivial systematic errors.

\section{Summary}
\label{sec-discussion}

We have shown that using the photometric redshift probability
distribution $p(z)$ greatly reduces photometric redshift systematic
errors, as compared to using a simple one-point estimate such as the
most probable redshift or the mean or median of $p(z)$.  Various
authors have made similar points previously, particularly
Fernandez-Soto \etal\ (2002), who wrote that ``this information [p(z)]
can and must be used in the calculation of any observable quantity
that makes use of the redshift.''  However, adoption of this practice
has been slow to nonexistent, even among authors who are aware of the
point, because it is cumbersome to track a full $p(z)$ for each
galaxy.  We have shown that a very simple modification to photometric
redshift codes, namely choosing a Monte Carlo sample from the $p(z)$,
produces a single number for each galaxy which greatly reduces the
systematic errors compared to using any other one-point estimate such
as the mean, median, or mode of $p(z)$.  In contrast to approaches
which simply reject galaxies which {\it could} be outliers, this
method can make use of every galaxy in a survey.  We have shown that
this method results in substantial improvements in a flagship
application, estimating dark energy parameters from weak lensing, at
no cost to the survey.

\end{document}